\documentclass[preprint,samsmath,amssymb]{revtex4}
\usepackage{graphicx}
\usepackage{epstopdf}
\usepackage{amsmath}

\begin{document}

\title{
A fully solvable equilibrium self-assembly process: \\ fine tuning the clusters size and the connectivity in patchy particle systems}

\author{ Emanuela Bianchi} 
\affiliation{ {Dipartimento di Fisica and  INFM-CNR-SMC, Universit\`a di Roma {\em La Sapienza}, Piazzale A. Moro 2, 00185 Roma, Italy} }
 \author{Emilia La Nave} 
\affiliation{ {Dipartimento di Fisica and  INFM-CNR-SOFT, Universit\`a di Roma {\em La Sapienza}, Piazzale A. Moro 2, 00185 Roma, Italy} } 
\author{Piero Tartaglia} 
\affiliation{ {Dipartimento di Fisica and  INFM-CNR-SMC, Universit\`a di Roma {\em La Sapienza}, Piazzale A. Moro 2, 00185 Roma, Italy} } 
\author{  Francesco Sciortino} 
\affiliation{ {Dipartimento di Fisica and  INFM-CNR-SOFT, Universit\`a di Roma {\em La Sapienza}, Piazzale A. Moro 2, 00185 Roma, Italy} }

\begin{abstract}
Self-assembly is the mechanism that controls the formation of well defined structures from disordered pre-existing parts.  Despite the importance of self-assembly  as a manufacturing method  and the increasingly large number of experimental realizations of complex self-assembled nano  aggregates, theoretical predictions are lagging behind.  Here we show that  for  a non-trivial self-assembly phenomenon, originating branched loop-less clusters, it is possible to derive a  fully predictive {\it parameter-free} theory of equilibrium self-assembly  by combining the Wertheim theory for associating liquids  with the Flory-Stockmayer approach for chemical gelation.  
\end{abstract}

\maketitle

 Intermolecular self-assembly is the ability of molecules to form supramolecular assemblies~\cite{pnas} as well as  a  manufacturing method used to construct  aggregate at the nano or micro scale, by proper design of the constituent molecules.   In the self-assembly bottom-up paradigm, the final (desired) structure is 'encoded' in the shape and properties of the designed building blocks. Realizations of complex self-assembled nano  aggregates~\cite{Kotov,Blaad06,XiaJACS}  have been guided by intuition and sophisticated experimental techniques. A  full comprehension of the self-assembly process requires the ability to predict the structures  (and their relative abundance)  which will be observed in equilibrium, as a function of 
$T$ and density $\rho$, starting from the knowledge of the 
inter-particle interaction potential. Such a request is very much akin to the one that has guided the development of the physics of liquids in the last decades.
 Differently from the simple liquid case, 
self-assembly is characterized by a very strong inter-particle attraction
(significantly larger than the thermal energy $k_BT$) and by the
fact that the interaction geometry is far from being spherical.
The leading "bonding" interaction may indeed be localized in 
a specific part of the particle surface (patchy interactions~\cite{Zhang_03}), it may be active only in the presence of a specific complementary group (lock-and-key interactions, very often encountered  in biological self-assembly~\cite{vecchio,Nanovirus,Zhang_04,Workum_06}) or it may be strongly dependent on the particle orientation~\cite{Glotz_Solomon_natmat}.
The presence of strong and patchy interactions poses
significant challenges to a parameter-free description of
self-assembly.  Only
 equilibrium chain polymerization,  the simplest self-assembly process which takes place  when bi-functional particles   self-assemble into chains of variable length, can be considered to be  sufficiently established~\cite {Greer02,rouault,Milchev98,Dudo_04,Kindt,M2}.

In this  article we show that  for systems with a small average functionality (but larger than two) it is possible to provide  a parameter-free full description of the self-assembly process. 
We study theoretically and numerically one of the simplest, but not trivial self-assembly process, namely a binary mixture of particles with two and three attractive sites.
The presence of  three  (or more) -functional  particles  --- which act as branching points in the self-assembled clusters --- introduces two important phenomena which are missing in equilibrium chain polymerization: a percolation transition, where a spanning cluster appears, and a region of thermodynamic instability, the analog of a gas-liquid phase separation. 
More explicitly, we investigate a binary mixture composed by $N_2=5670$ bi-functional particles and $N_3=330$ three-functional ones. The resulting average number of sticky spots per particle, i.e. the average  functionality, is $\langle M\rangle=\frac{2N_2+3N_3}{N_2+N_3}$.  
Particles are modeled as hard-spheres of diameter $\sigma$, whose surface is decorated by
two (or three)  bonding sites at fixed locations. Sites on different particles interact via a square-well potential $V_{SW}$ of depth $u_0$ and attraction range $\delta=0.119 \sigma$. 
More precisely, the  interaction potential $V({\bf 1,2})$  between particles {\bf 1} and  {\bf 2}  is
\begin{equation}
V({\bf 1,2})=V_{HS}({\bf r_{12}})+\sum_{i=1,n_1}\sum_{j=1,n_2} V_{SW}({\mathbf r}^{_{ij}}_{_{12}})
\end{equation}
where  $V_{HS}$ is the hard-sphere potential  and  ${\bf r_{12}}$ and ${\mathbf r}^{_{ij}}_{_{12}}$  are respectively the vectors joining the particle-particle centers and the site-site (on different particles) locations; $n_i$ indicates the number of sites of particle $i$. 
Sites are located on the poles in the case of bi-functional particles and equidistant on the equator in the case of three-functional particles.  With this site geometry, the smallest possible bonded ring requires six three-functional particles, while a  bonded ring of only bi-functional units requires  $\approx 50$ particles. The well width $\delta$ is chosen to ensure that each site, due to steric effect, is
engaged at most in one interaction. Distances are measured in units of $\sigma$.
Temperature is measured in units of the potential depth (i.e. Boltzmann constant $k_B=1$). 
In the studied model,   bonding is properly defined: two particles are bonded when their pair interaction energy is -$u_0$. This means that the potential energy of the system is proportional to the number of bonds. The lowest energy state of the system (the ground state energy)  coincides with configurations in which all bonds are formed, i.e $E_{gs}=-u_0\frac{2N_2+3N_3}{2}$. As a result,  the bond probability, $p_b$, can be precisely calculated as the ratio of the potential energy $E$  and $E_{gs}$. Pairs of  bonded particles are assumed to belong to the same cluster. 

We have performed standard Monte Carlo (MC)  Metropolis simulations at   several $T$ and $\rho$ (more than 70 different state points). An MC step is defined as an attempted move per particle.  A  move is defined as a displacement of a randomly selected particle in each direction of a random quantity distributed uniformly between $\pm 0.05~\sigma$ and a rotation around a random axis of a random angle uniformly  distributed between $\pm 0.1$ radiant.   Equilibration was monitored via the evolution of the potential energy (a direct measure of the number of bonds in the system). 
 Equilibration at the lowest $T$ required up to $10^9$ MC steps (6 months of simulation time). Indeed, at $k_BT/u_0=0.05$,
a bond breaks in average every $5~10^8$ attempts. 

 The extremely long  Monte Carlo simulations  provide a numerically "exact" description of 
the equilibrium self-assembly process in this model.   On cooling particles aggregate in open larger and larger clusters which then coalesce  into a spanning  structure, eventually incorporating all particles.  A pictorial representation of the structure of the system on varying $T$ (both below, at and above percolation) is shown in  Fig.~\ref{fig:picture}.  
Clusters are composed by  rather stiff chains of bi-functional particles cross-linked by the
three-functional ones.

To develop a parameter-free  theoretical description of the cluster formation  we combine  Wertheim (W) theory~\cite{Werth1,Hansennew,Jack_88} for estimating the $T$ and $\rho$ dependence of  $p_b$ with the Flory-Stockmayer (FS) predictions for the cluster size distribution in chemical gelation, providing   a detailed and predictive theory of reversible self-assembly.

The Wertheim theory, developed  back in 1980  for
describing the free energy of molecules with fixed valence (associating liquids), can be 
transferred to particles with a small number of patchy interacting sites. 
The main assumption in the theory is that molecules (or particles) cluster in open structures
without closed bond loops. Such a condition, as we will show in the following, can be realized with patchy particles when the average functionality is small and the chains of bi-functional  particles are not significantly flexible. The Wertheim theory
 predicts that  $p_b$ can be calculated from the chemical equilibrium between two non-bonded sites forming a bonded pair. For the present model, such relation reads:

\begin{equation}
\frac{p_b}{(1-p_b)^2}=\langle M\rangle \rho \Delta
\label{eq:pb}
\end{equation}
where 
$\Delta= 4 \pi \displaystyle \int{g_{HS}(r_{12})\langle f(12)\rangle _{\omega_{1},\omega_{2}} r_{12}^{2} d r_{12}}.$
Here $g_{HS}(12)$ is the reference hard-sphere radial distribution function, $f(12) \equiv  e^{-V_{SW}({\mathbf r}_{12}^{ij})/k_B T}-1$ 
is the Mayer $f$-function between two arbitrary sites $i$ and $j$, and $\langle f(12)\rangle _{\omega_{1},\omega_{2}}$ represents an angle average over all orientations of the two particles  at fixed relative distance $r_{12}$~\cite{Werth5}. The comparison between the theoretical predictions and the "exact" numerical data for the  $T$ and $\rho$ dependence of $p_b$  are shown  in Fig.~\ref{fig:pb}.  Data show clearly that
the theory is able to predict  precisely  $p_b$ (or equivalently the system potential energy) in a wide $T$ and $\rho$ range. At low $T$,  $p_b \rightarrow 1$, and the system approaches a fully bonded disordered (ground state) configuration. 

To derive information on the structure of the system and the connectivity of the aggregates
we connect the W and the FS theories. Indeed, the hypothesis of absence of closed bonding loops is at the basis of both theoretical approaches. The W prediction for $p_b$ can thus be consistently used in connection with the FS approach~\cite{flory} to predict the $T$ and $\rho$ dependence of the cluster size distributions. In the present case, the number of clusters (per unit volume) containing $l$ bi-functional particles and $n$ three-funtional ones can be written~\cite{flory} as
\begin{eqnarray}\label{eq:rholm}
\rho_{nl} & = &  \rho_3  \frac{(1-p_b)^2}{p_{3} p_b }  [p_{3} p_b  (1-p_b)] ^n [p_{2} p_b ]^l  w_{nl}\\
\nonumber
w_{nl} & = & 3  \frac{(l+3n-n)!}{l! n! (n+2)!}
\end{eqnarray}
where $p_{3} \equiv  3 N_3/(2N_2+3N_3)$ and $p_{2}=1-p_{3}$ are the probabilities that  a randomly chosen site  belongs to a three-functional or to a bi-functional particle,  $p_b$ is  given by Eq.~\ref{eq:pb}, and $w_{nl}$ is a combinatorial contribution~\cite{flory}. Distributions are normalized  in such a way that $\sum_{ln,l+n>0} (l+n) \rho_{nl}=\rho_2+\rho_3$.  As shown  in Fig.~\ref{fig:csd}-(top), on decreasing $T$, the  $\rho_{nl}$ distribution becomes wider and wider and develops a power-law tail with exponent -2.5, characteristic of loop-less percolation~\cite{colby}. On further decreasing $T$, the distribution of finite size clusters progressively shrinks, since most of the particles attach themselves to the infinite cluster.   Data show that
Eq.~\ref{eq:rholm}, with no fitting parameters, predicts extremely well the numerical distributions  at all state points, both above and below percolation.

The three-functional particles act as branching points of the network formed by long chains of two-functional  particles.
Visualizing the  structure of the system in term of chains of two-coordinated particles providing a link between the three-coordinates ones, it is possible to predict the number of finite size clusters composed of $n$ three-functional units, irrespective of the number of bifunctional units. The system can thus be considered as a one-component fluid of three-functional particles forming clusters, in which the bonding distance between the three-functional particles is given by the length of the chains formed by the bi-functional units. Following again FS , it is possible to   predict  the $p_b$ value at which the systems develops a percolating structure:  when $p_b \geq p_b^c  \equiv \frac{1}{1+p_{3}}=0.9256$, an infinite cluster is present in the system.   The percolation line is thus the locus of points in the phase diagram such that $p_b(T,\rho) = p_b^c$,
with $p_b(T,\rho)$ given in Eq.\ref{eq:pb}.  Along the percolation locus, the product $\rho \Delta $   ($\rho e^{1/T}$ at low $T$)  is constant (from  Eq.\ref{eq:pb}).

 It is also possible to predict  the length  distribution of chains of bi-functional particles.  The number of chains (per unit volume) of length $l$, normalized in such a way that  $\sum_{l=1}^{\infty} l \rho_{l}=\rho_2$,  is
\begin{equation}
\rho_{l}=\rho_2  (1-p_{2} p_b)^2  (p_{2} p_b)^{l-1}
 \end{equation}
The $\rho_l$ distribution is thus always exponential. At low $T$, when $p_b \rightarrow 1$,  the distribution becomes  controlled only by the relative fraction of two-functional particles, providing a method for tuning the porosity of the fully connected percolating structure via the relative composition of the binary mixture. Indeed, when $p_b \rightarrow 1$ , the average distance $\bar l$ between branching points in the network becomes  only a function of $p_2$, i.e.  $\bar l =1/(1- p_{2})$.
Fig.~\ref{fig:csd}-(bottom)  shows that also the $\rho_l$ distribution  is perfectly  described by the  combined W-FS theory.


In the framework of FS approach it is also possible to evaluate the number density of clusters $\rho_c \equiv  \sum_{nl} \rho_{nl}$ as a function of  $p_b$, irrespectively of the cluster size. 
Below percolation, in the absence of bonding loops, the relation between $\rho_c$ and $p_b$ is linear, since each added bond decreases the number of clusters by one. 
Above percolation the relation crosses to a non-linear behavior, so that the number of clusters becomes one when $p_b=1$.  Within the FS theory, $\rho_c$ can be calculate for all $p_b$ values, assuming that finite clusters do not contain closed loops. 
As shown in Fig.~\ref{fig:ncl}, the simulation data conform perfectly to the theoretical expectation both  below and above percolation. This  suggests  that, when the average functionality is small and the chains of bi-functional 
particles have a large persistence length, bonding loops in finite size clusters can be neglected.   This agreement, which covers the entire range of  $p_b$ values, implies  that  closed loops of bonds are statistically less favored than the corresponding open  structure. The relative statistical weight  results from the competition   between the energy of forming the extra additional bond and the reduction of entropy   associated to the closure of the loop. Hence, we interpret 
the absence of closed loops as resulting from  the large configurational entropy
 of the long bi-functional  chains.

To further check the quality of the theory,  we numerically evaluate the connectivity properties of each studied state point, searching for the presence of clusters which are infinite under periodic boundary conditions.   A state point is considered percolating   when,
accounting for periodic boundary conditions, an infinite cluster is
present in more than  50$\%$ configurations.  The resulting partitioning of the state points into percolating and non-percolating ones (see Fig.~\ref{fig:phase}) is fully consistent with the theoretical prediction of the percolation line $p_b(T,\rho)=p_b^c$. 
The theory also predict a line of  constant volume specific heat $C_V$ maxima (also observed in chain polymerization~\cite{Greer02,douglas}),
provided by the inflection point in the $p_b$ vs. $T$ curves (Fig.~\ref{fig:pb}),  which also agrees very well with the simulation results.  The line of  $C_V$ extrema  is also shown in Fig.~\ref{fig:phase}. The presence of a maximum in $C_V$  is a characteristic of bond-driven assembly and the locus of maxima in the  $T$-$\rho$ plane is one of the  precursors of the self-assembly process for low functionality particles. 

The W theory  predicts a liquid-gas phase separation at small $\rho$ for any non-vanishing amount of branching point~\cite{Sear_99,bian}. According to the theory, at low $T$ and sufficiently small $\rho$ the system phase separates into two phases of different density and connectivity.  The theoretical spinodal curve, the line separating the stable (or metastable) state points  from the unstable ones, is defined as the locus of points such that the volume $V$ derivative of the pressure $P$ vanishes, i.e. $(\partial P/\partial V)_{T}=0$. For the present model, it is located below the percolation line and the two lines merge asymptotically  for $T \rightarrow 0$ and $\rho \rightarrow 0$ as shown in Fig.~\ref{fig:phase}.   The analysis of the numerical configurations  for the two investigated state points which happened to be located
inside the spinodal are indeed characterized by a bimodal distribution of the density fluctuations and a very large value of the small-angle structure factor, indicating a phase-separated structure. This confirms that, in the absence of bonded rings, also the region of liquid-gas instability can be predicted using the W theory.  The role of bond-rings on the thermodynamic of thermoreversible gels has been recently discussed in Ref.~\cite{Kindt}.

To summarize, this work provides the first fully solvable example of  an equilibrium self-assembly process which goes beyond the equilibrium chain polymerization, complementing recent numerical studies and attempts at direct comparison between theory and simulation in related systems~\cite{Kumar,Kindt,Kindtrev}. The presence of a non negligible number of three-valent  particles  brings in a percolation phenomenon and a gas-liquid instability. These two loci are located in the region of $T$ below the  $C_V$ maximum, suggesting a cascade of phenomena characterizing the self-assembly process in systems with low $ \langle M \rangle $: onset of bonding ($C_V$ max), percolation and, eventually at low $\rho$, phase separation.  
The resulting phase diagram bears strong similarities with the ones
 discussed in previous mean field~\cite{csk,Safranrev}  and numerical~\cite{Kumar,Zacca1,Kindt} studies of thermoreversible gelation.

The possibility of theoretically describing the self-assembly process  in the present model, designed to minimize the closed-bond loops effects, provides a benchmark for testing novel approaches and approximations accounting for  more general cases. The quality of the predictions makes it possible to exploit a fine tuning of the percolation properties. 
Using the recently synthesized colloidal particles~\cite{Manoh_03,Cho_05} or functionalizing colloids~\cite{Mirkin_96, Kiang,Stellacci} it will be possible, by choosing the appropriate $\rho$ and $T$, 
 to generate materials with specific connectivity lengths, gels with
desired porosity or build on the fractal nature of the self-assembled clusters. The material structure can be made permanent, if needed, by freezing the bonding pattern with  a fast cooling process.

Finally, $p_b$ appears to be the effective parameter controlling the structure of the system, since  configurations at different $\rho$ and $T$ but equal $p_b$ are characterized by the same distribution of cluster sizes. Configurations only differ in the relative distance between these clusters and, more interesting, on the lifetime of the resulting bonds, $\tau \sim e^{u_0/k_BT}$. It is thus in principle possible to decouple the effects related to the persistence of the clusters as units and the dynamics of clusters. The possibility of  interpolating in a continuous and structural preserving way from chemical to physical gels may offer a unifying picture of these two apparently different arrest processes.

\clearpage
\section*{Figure Captions}

\begin{figure}[h]
\caption{ Representation of the studied system for three different temperatures (in units of $u_0$) at density $\rho \sigma^3=0.04$. Equally colored particles belong to the same cluster. The pictures show the structure of the system below percolation (left, $p_b < p_b^c$), where particles are aggregated in small finite clusters, at percolation (center, $p_b \approx p_b^c$), where a spanning cluster first appears, and
well above percolation (right, $p_b \gg p_b^c$), where most all of the particles belong to the infinite cluster.  The mesh size of the fully connected system  --- provided by the average length of the chains connecting the three-functional branching points --- can be finely controlled by the relative composition of the mixture.}
\label{fig:picture}
\end{figure}


\begin{figure}[h]
\caption{ Temperature  (in units of $u_0$) and density dependence of the bond probability $p_b$. Points are simulation results based on Monte Carlo simulations for nine different densities and eight  different temperatures. Lines are parameter-free predictions based on the Wertheim theory. Note that at low $T$ the system reaches a fully bonded configuration. }
\label{fig:pb}
\end{figure}


\begin{figure}[h]
    \caption{Cluster size distribution for $\rho \sigma^3=0.04$ for some of the
   investigated $T$. Points are simulation data and lines are the corresponding theoretical curves. (top) The number of finite clusters (per unit volume) containing $l$ bi-functional units plus $n$ three-functional ones.     (bottom) The number of chains of length $l$ (per unit volume) composed of bi-bunctional particles.  At low $T$, the distribution approaches its limiting (concentration controlled) value  $\rho_l=\rho_2 p_3^2 p_2^{l-1}$, shown by the dashed line.}
   \label{fig:csd}
\end{figure}


\begin{figure}[h]
\caption{ Relation between the number of finite-size clusters  $N_c$ (irrespective of their size) and the bond probability. Symbols: simulation results  for all studied densities. Solid line: Flory-Stockmayer predictions (both below and above percolation). 
Below percolation, in the absence of bonding loops,  $N_c$  is given by the difference between the total number of simulated particles $N=6000$ and  the number of bonds $N_b$, since each added bond decreases  $N_c$ by one, i.e. $N_c=N-N_b= N - p_b \frac{2N_2+3N_3}{2}$.  This linear relation (dashed line) can be extended above percolation (dotted line), but  never beyond the point where $N_c<1$.  The inset enlarges the region of large $p_b$ values, to provide evidence that the FS approach is valid over the entire $p_b$ range. }
\label{fig:ncl}
\end{figure}


\begin{figure}[h]
\caption{ Phase-diagram of the studied model.
Lines are theoretical predictions: the solid  (green) line is the spinodal curve obtained from the Wertheim equation of state, i.e. $(\partial P/\partial V)_{T}=0$;  the  dashed (magenta) line is the locus of points such that  the specific heat is maximum, i.e. $(\partial C_V/\partial T)_V=0$; the dotted (red) line is the percolation locus  $p_b(\rho,T)=p_b^c$ . Points are simulation results: the (black) triangles are the non-percolating equilibrium state points and the (red) squares are the equilibrium percolating state points.  Circles are (non-equilibrium) state points characterized by a bimodal distribution of the density fluctuations, indicating a phase-separated structure. 
}
\label{fig:phase}
\end{figure}


\section*{Acknowledgements}
We acknowledge support from  MIUR-Prin and MCRTN-CT-2003-504712.

\clearpage

\setcounter{figure}{0}

\begin{figure}[h]
\includegraphics[width=19.0 cm, clip=true]{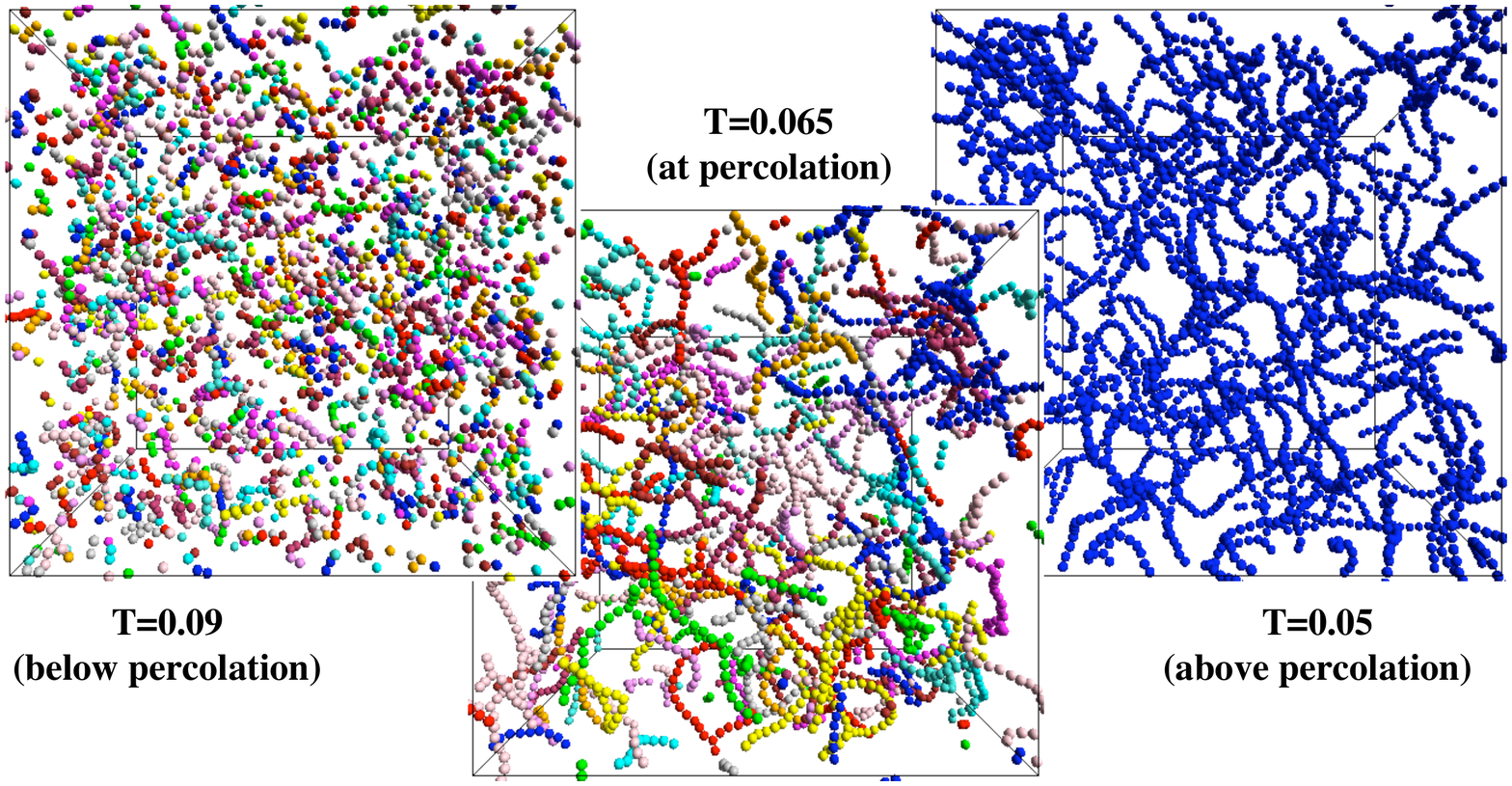}
\caption{
}
\label{fig:picture}
\end{figure}

\clearpage

\begin{figure}[h]
\includegraphics[width=15.0 cm, clip=true]{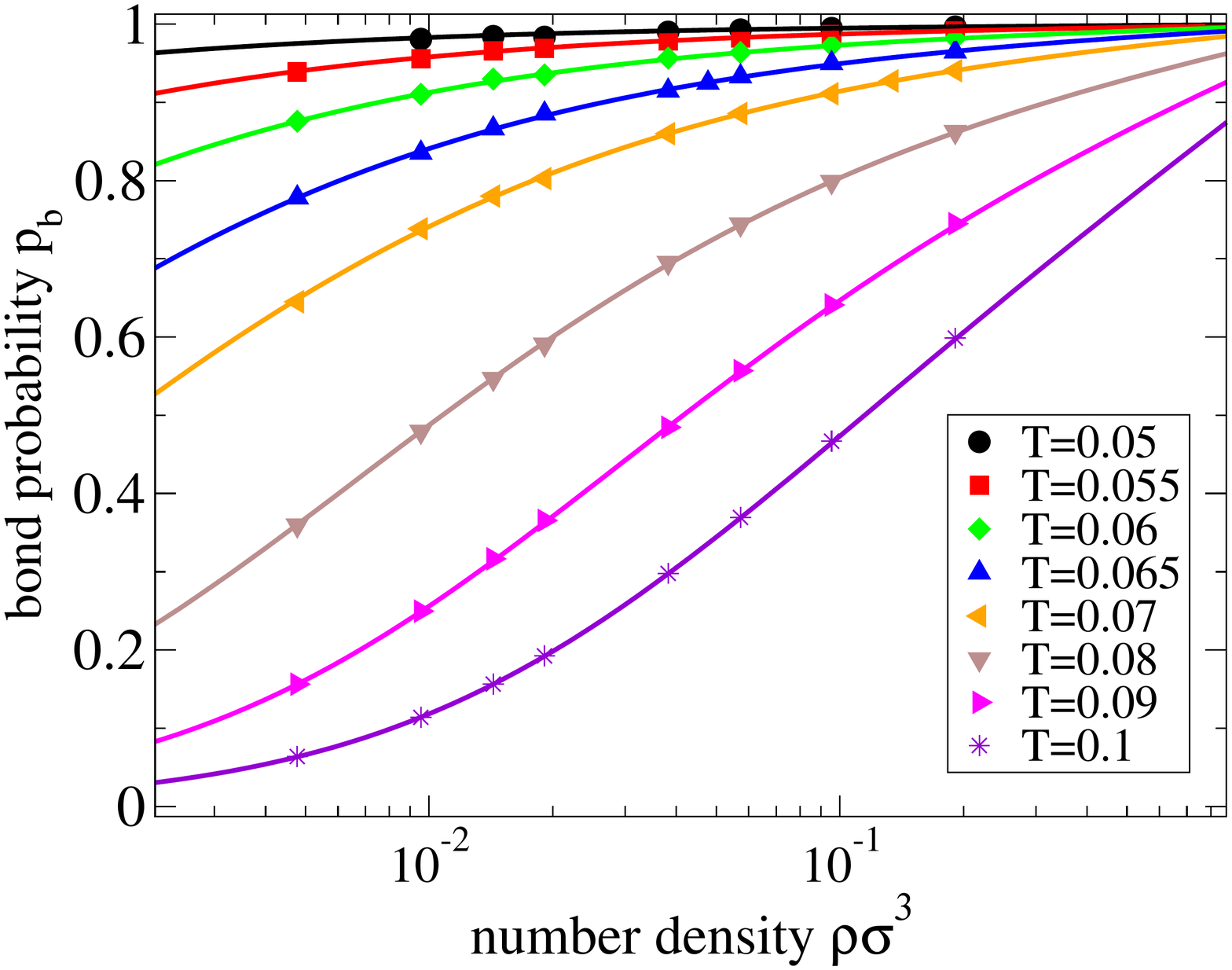}
\caption{ 
}
\label{fig:pb}
\end{figure}

\clearpage

\begin{figure}[h]
 \includegraphics[width=15.0 cm, clip=true]{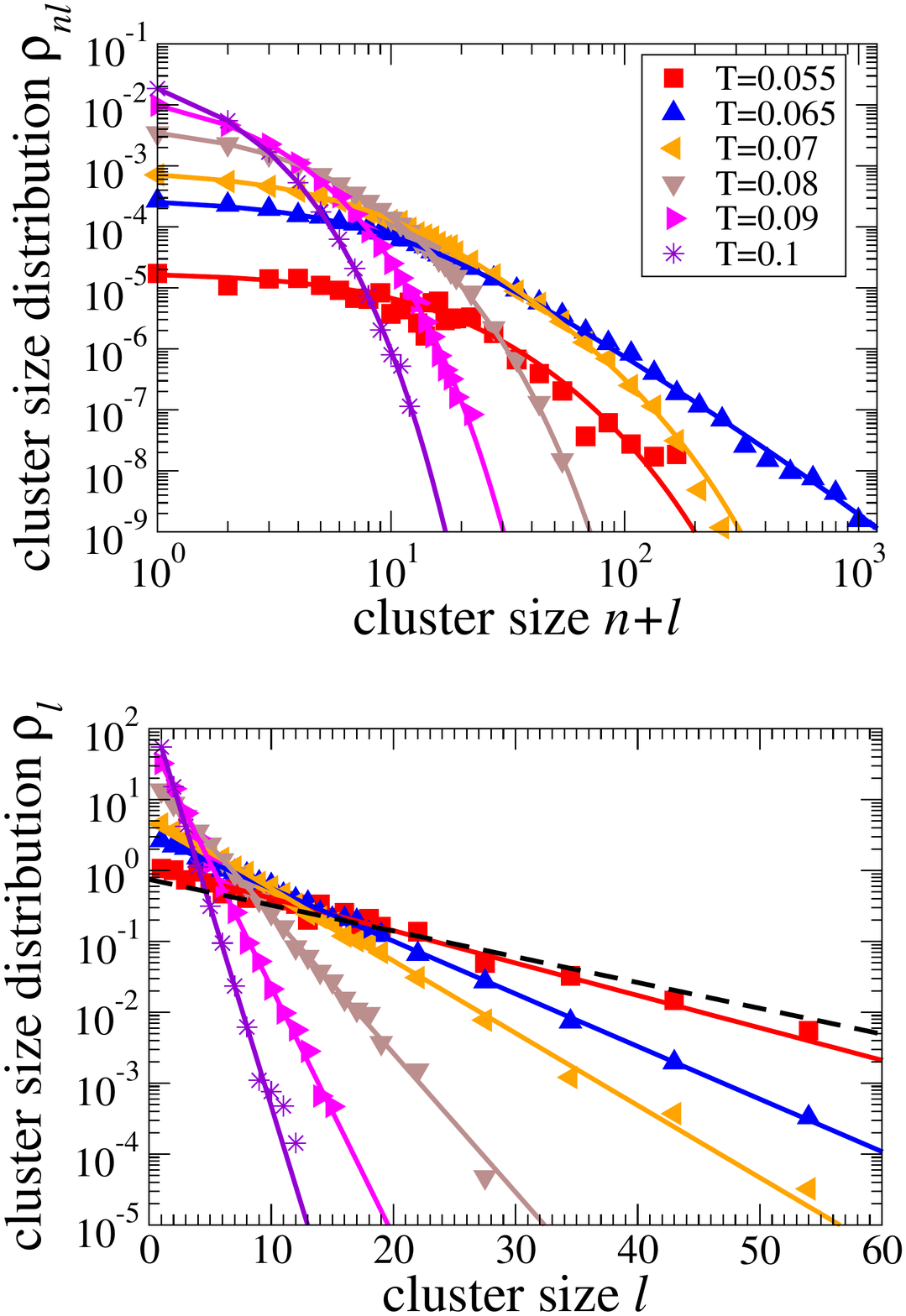}
 \caption{
}
   \label{fig:csd}
\end{figure}

\clearpage

\begin{figure}[h]
\includegraphics[width=15.0 cm, clip=true]{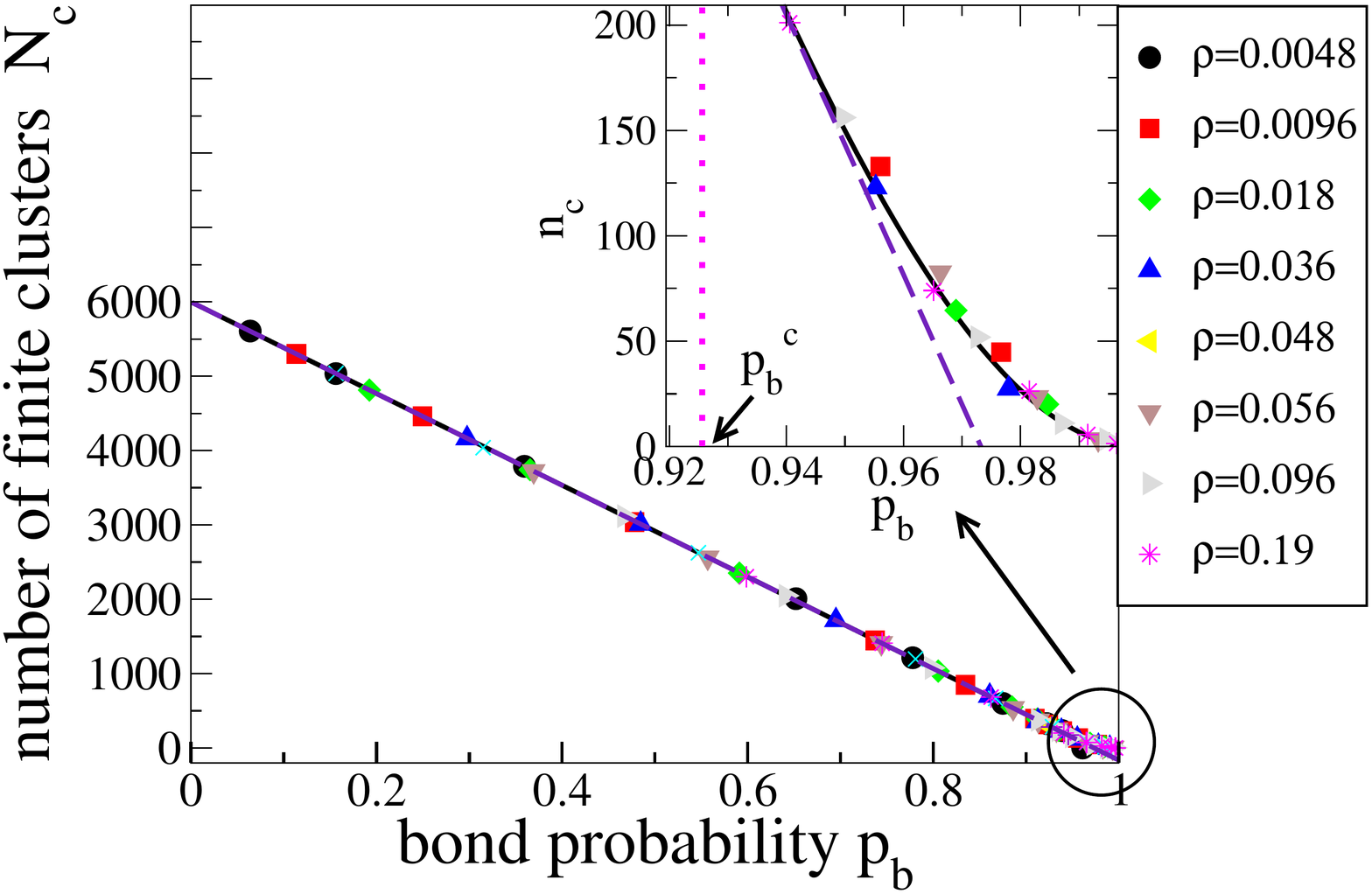}
\caption{
 }
\label{fig:ncl}
\end{figure}

\clearpage

\begin{figure}[h]
\includegraphics[width=15.0 cm, clip=true]{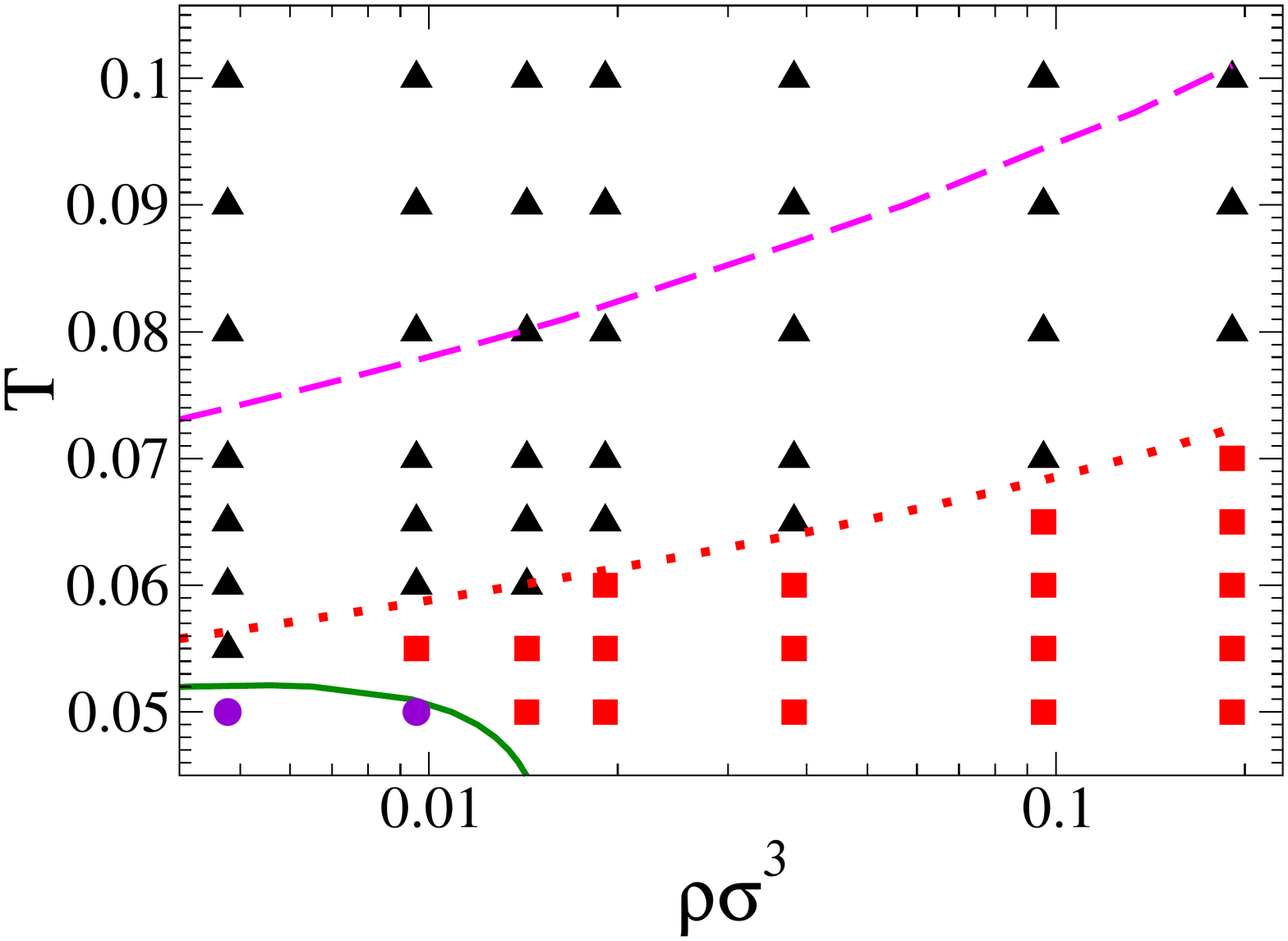}
\caption{ 
}
\label{fig:phase}
\end{figure}

\end{document}